\documentclass[10pt]{article}

\usepackage{amsmath}
\usepackage{amssymb}
\def\beq{\begin{equation}}
\def\eeq{\end{equation}}

\newcommand{\ed}{\end{document}}

\begin{document}
\title{A brief review on the Problem of Divergence in Krein Space Quantization}

\author{Farrin Payandeh$^{1,2}$\thanks{e-mail:
f$\_$payandeh@pnu.ac.ir} , Zahra Gh. Moghaddam$^3$\thanks{e-mail:
zahra$\_$ghmoghaddam@yahoo.com} , Mohsen Fathi$^3$\thanks{e-mail:
mohsen.fathi@gmail.com}}

\maketitle   \centerline{\it $^{1}$Department of Physics, Payame
Noor University, PO BOX 19395-3697, Tehran, Iran}
\centerline{\it$^2$Department of Physics, Amirkabir University of
Technology, Tehran 15914 Iran} \centerline{\it $^3$Department of
Physics, Islamic Azad University, Central Tehran Branch, Tehran,
Iran}
\begin{abstract}

In this paper we have a brief review on the problem of divergence
in quantum field theory and its elimination using the method of
Krein space quantization. In this method, the auxiliary negative
frequency states have been utilized, the modes of which do not
interact with the physical states and are not affected by the
physical boundary conditions. It is remarkable that Krein space
quantization is similar to Pauli-Villars regularization, so we can
call it the ''Krein regularization''. Considering the QED in Krein
space quantization, it could be shown that the theory is
automatically regularized. Calculation of the three primitive
divergent integrals, the vacuum polarization, electron self energy
and vertex function using Krein space method leads to finite
values, since the infrared and ultraviolet divergencies do not
appear. For another example, the Casimir stress on a spherical
shell in de Sitter spacetime for a massless scalar field could be
calculated using Krein space quantization.

\end{abstract}

\maketitle

\section{Introduction}

The historical background of Krein space quantization goes back to
the covariant quantization of minimally coupled scaler field in de
Sitter spacetime. It has been shown that the linear quantum
gravity in the background field method is perturbatively
non-renormalizable and also there appears an infrared divergence.
This infrared divergence does not manifest itself in the quadratic
part of the effective action in the one-loop approximation. This
means that the pathological behavior of the graviton propagator
may be  gauge dependent and so should not appear in an effective
way as a physical quantity \cite{anilto}. The infrared divergence
which appears in the linear gravity in de Sitter space is the same
as the minimally coupled scalar field in de Sitter space
\cite{gagarota,derotata}. It is shown that one can not construct a
covariant quantization of the minimally coupled scalar field with
only positive norm states \cite{al}. It has been proved that the
use of the two sets of solutions (positive and negative norms
states) is an unavoidable feature if one wants to preserve
causality (locality), covariance and elimination of the infrared
divergence in quantum field theory for the minimally coupled
scalar field in de Sitter space \cite{gareta,ta3}, {\it i.e.}
Krein space quantization.

The singular behavior of Green function at short relative
distances (ultraviolet divergence) or in the large relative
distances (infrared divergence) leads to main divergences in the
quantum field theory. It was conjectured that quantum metric
fluctuations might smear out the singularities of Green functions
on the light cone, but it does not remove other ultraviolet
divergences \cite{for2}. However, it has been shown that
quantization in Krein space removes all ultraviolet divergences of
quantum field theory (QFT) except the light cone singularity
\cite{ta4}.  By using the Krein space quantization and the quantum
metric fluctuations in the linear approximation, it has been shown
that the infinities in the Green function are disappeared
\cite{for2,rota}.

Quantization in Krein space instead of Hilbert space has some
interesting features. For example in this method, the vacuum
energy becomes zero naturally, so the normal ordering would not be
necessary \cite{gareta,ta4}. The auxiliary negative norm states,
which are used in the Krein space quantization, play the role of
regularization in the theory. Since Krein space quantization is
similar to Pauli-Villars regularization, so it could be called the
''Krein regularization''. In the the Pauli-Villars regularization
the particles with mass $M$ and negative norms are added to the
theory. In Krein space the negative norm states with mass $m$ are
also added the same as the physical mass particle. So, considering
the QED in Krein space quantization, results in an automatically
regularized theory. Calculation of the three primitive divergent
integrals, the vacuum polarization, electron self energy and
vertex function using Krein space method leads to finite values,
since the infrared and ultraviolet divergencies do not appear
\cite{Boson, AIP, QFT, Mollerm, }.

 One of the interesting features of quantization in Krein space
instead of Hilbert space is that the vacuum energy becomes zero
naturally, so the normal ordering would not be necessary
\cite{gareta,ta4}.

For another example, the Casimir effect could be investigated
using Krein space method, too. Since the original work by Casimir
in 1948 \cite{Casi48} many theoretical and experimental works have
been done on this problem. The Casimir effect is a small
attractive force acting between two parallel uncharged conducting
plates and it is regarded as one of the most striking
manifestation of vacuum fluctuations in quantum field theory. It
is due to the quantum vacuum fluctuation of the field operator
between two parallel plates. In other words, the Casimir effect
can be viewed as the polarization of the vacuum by boundary
conditions or geometry. The presence of reflecting boundaries
alters the zero-point modes of a quantized field, and results in
the shifts in the vacuum expectation values of quantities
quadratic in the field, such as the energy density and stresses.

Applying the unphysical negative frequency states and defining the
field operator in Krein space, the Casimir stress on a spherical
shell in de Sitter spacetime for a massless scalar field could be
calculated using Krein space quantization, yielding the standard
result obtained \cite{Naseri,Casimir2}.

\section{A brief review on Krein space quantization}

In Krein space the quantum scalar field is defined as follows
\cite{gareta,ta3}:

$$ \phi(x)=\frac{1}{\sqrt 2}[\phi_p(x)+\phi_n(x)], $$

where
$$
\phi_p(x)=\int d^3\vec k [a(\vec k)u_p(k,x)+a^{\dag}(\vec
k)u_p^*(k,x)],$$ $$ \phi_n(x)=\int d^3\vec k [b(\vec
k)u_n(k,x)+b^{\dag}(\vec k)u_n^*(k,x)].
$$

$a(\vec k)$ and $b(\vec k)$ are two independent operators and
$$
u_p(k,x)=\frac{e^{i\vec k.\vec x-iwt}}{\sqrt{(2\pi)^32w}}
=\frac{e^{-ik.x}}{\sqrt{(2\pi)^32w}},\;\;u_n(k,x)=\frac{e^{-i\vec
k.\vec x+iwt}}{\sqrt{(2\pi)^32w}}
=\frac{e^{ik.x}}{\sqrt{(2\pi)^32w}},
$$

where $ w(\vec k)=k^0=(\vec k .\vec k+m^2)^{\frac{1}{2}} \geq 0$.
The positive mode $\phi_p$ is the scalar field operator as was
used in the usual QFT and $\phi_n$ plays the role of the
regularization field. The time-ordered product is defined as:
$$
iG_T(x,x')=<0\mid T\phi(x)\phi(x') \mid 0>=\Re G_F(x,x'),
$$

where $G_F(x,x')$ is the Feynman Green function.

As we know, the origin of divergences in standard quantum field
theory lies in the singularity of the Green's function. The
divergence appears in the imaginary part of the Feynman
propagator, and the real part is convergent \cite{QFT}:
$$
G^P_F(x,x^\prime)=-\frac{1}{8\pi}\delta(\sigma_0)+\frac{m^2}{8\pi}\
\theta(\sigma_0)[\frac{J_1(\sqrt{2m^2\sigma_0})
-iN_1(\sqrt{2m^2\sigma_0})}{\sqrt{2m^2\sigma_0}}]-
\frac{im^2}{4\pi^2}\
\theta(-\sigma_0)\frac{K_1(\sqrt{2m^2(-\sigma_0)})}{\sqrt{2m^2(-\sigma_0)}}
$$

Consideration of negative frequency states removes singularity of
the Green function with exception of delta function singularity:

$$
G_T(x,x^\prime)=-\frac{1}{8\pi}\delta(\sigma_0)+
\frac{m^2}{8\pi}\theta(\sigma_0)\frac{J_1(\sqrt{2m^2\sigma_0})}{\sqrt{2m^2\sigma_0}},
\,\ \,\ \sigma_0\geq0
$$

However, considering the quantum metric fluctuations removes the
latter singularity:

\begin{equation}
<G_T(x,x^\prime)>=-\frac{1}{8\pi}\sqrt{\frac{\pi}{2<\sigma_1^2>}}\exp(-\frac{\sigma_0^2}{2<\sigma_1^2>})
+\frac{m^2}{8\pi}\theta(\sigma_0)\frac{J_1(\sqrt{2m^2\sigma_0})}{\sqrt{2m^2\sigma_0}}.
\label{Green}
\end{equation}
\noindent where $<\sigma_1^2>$ is related to the density of
gravitons. When $\sigma_0=0$, due to the metric quantum
fluctuation $<\sigma_1^2>\neq0$, and we have
$$
<G_T(0)>=-\frac{1}{8\pi}\sqrt{\frac{\pi}{2<\sigma_1^2>}}+\frac{m^2}{16\pi}.
$$

By using the Fourier transformation, we obtain \cite{Takook2}

$$
<\widetilde{G}_T(p)>=\widetilde{G}_T(p)+
PP\frac{m^2}{p^2(p^2-m^2)}
$$

However, in the one-loop approximation, the contribution of delta
function is negligible and the Green function in Krein space
quantization appearing in the transition amplitude is

$$
<\widetilde{G}_T(p)>\mid_{one-loop}\equiv{\widetilde{G}_T(p)\mid_{one-loop}}\equiv
PP\frac{m^2}{p^2(p^2-m^2)}
$$

where $\tilde{G}_1(p)$ is the Fourier transformation of the first
part of the Green function $(1)$ and its explicit form is not
needed for our discussion here. In a previous paper, it has proved
that for the $\lambda\varphi^4$ theory in the one-loop
approximation, the Green function in Krein space quantization,
which appear in the s-channel contribution of transition
amplitude, is the second part of $(1)$ \cite{ta3}. That means in
this approximation, the contribution of the first part (i.e.
quantum metric fluctuation) is negligible. It is worth mentioning
that in order to improve the UV behavior in relativistic
higher-derivative correction theories, the second part of equation
(1) has been used by some authors \cite{ba,ho}. This part also
appears in the super-symmetry theory \cite{ka}.

The time-order product of the spinor field is:
$$
<S_T(x-x')>\equiv (i\not\partial+ m )<G_T(x,x')>
$$

And the time-ordered product propagator in the Feynman gauge for
the vector field in Krein space is given by:

$$
<D_{\mu\nu}^T(x,x^\prime)>=-\eta_{\mu\nu} <G_T(x,x^\prime)>.
$$

\section{Essential graphs of QED in Krein space quantization}

In the standard QED the divergent quantities are found in the
electron self-energy, the vacuum polarization and the vertex
graphs. In the standard QED, we have \cite{pesc}:

$$\Sigma _{Hi} (p) =\frac{e^2 }{8\pi ^2 }\left\{ \ln \left(
- \frac{\Lambda^2 }{m^2} \right)\left(2m - \frac{\not p}{2}
\right) +\left( 2m-\frac{3}{4} \not p\right) \right. $$
$$\left. { -
\frac{{\not p}} {2}\left[ {\frac{{m^4  - (p^2 )^2 }} {{(p^2 )^2
}}\ln \left( {1 - \frac{{p^2 }} {{m^2 }}} \right)} \right] +
 2m\left[ {\frac{{m^2  - p^2 }} {{p^2 }}\ln \left( {1 -
\frac{{p^2 }} {{m^2 }}} \right)} \right]} \right\}.  $$

and

$$
\Pi_{Hi} (k^2 ) =\frac{{e^2 }} {{12\pi ^2 }}\ln\left( {
\frac{{\Lambda^2 }} {{m^2 }}} \right) - \frac{{e^2 }}
{{2\pi^2}}\int\limits_0^1 {dx} (1 - x)x\ln \left(1 - x(1 -
x)\frac{{k^2 }} {{m^2 }}\right) \hfill.
$$

and

$$
F_{1}^{Hi} (q^2 )_{q^2  \to 0} =-\frac{{e^2 }} {{16\pi ^2
}}\ln\left( {  \frac{{\Lambda^2 }} {{m^2 }}} \right)- \frac{{e^2
q^2 }} {{12\pi^2 m^2 }}\left(\ln\frac{m} {\mu } - \frac{3}
{8}\right) .
$$

Calculating in Krein space, we get:

$$
\Sigma _{kr} (p) = \frac{{e^2 }} {{8\pi ^2 }}\left\{ {\ln \left( {
- \frac{{p^2 }} {{m^2 }}} \right)\left( {2m - \frac{{\not p}} {2}}
\right) - \frac{{\not p}} {2}\left( {\frac{{m^2 }} {{p^2 }}}
\right)} \right.\
$$

$$ \left. { - \frac{{\not p}} {2}\left[ {\frac{{m^4  - (p^2 )^2 }}
{{(p^2 )^2 }}\ln \left( {1 - \frac{{p^2 }} {{m^2 }}} \right)}
\right] +
 2m\left[ {\frac{{m^2  - p^2 }} {{p^2 }}\ln \left( {1 -
\frac{{p^2 }} {{m^2 }}} \right)} \right]} \right\}.
$$

and

$$
\Pi _{\mu \nu }^{kr} (k^2 ) =(k^2 g^{\mu \nu }  - k^\mu  k^\nu
)\Pi_{kr}(k^2
 ),
$$
where
$$
\Pi_{kr} (k^2 ) =  -\frac{{e^2 }} {{12\pi ^2 }}\ln\left( { -
\frac{{k^2 }} {{m^2 }}} \right)- \frac{{e^2 }} {{6\pi ^2
}}\frac{{k^2 }} {{m^2 }} - \frac{{e^2 }} {{2\pi^2}}\int\limits_0^1
{dx} (1 - x)x\ln \left(1 - x(1 - x)\frac{{k^2 }} {{m^2 }}\right)
\hfill .
$$

and

$$
\Lambda_{kr} ^\mu  (p',p) = \frac{{e^2 }} {8\pi}\int \frac{d^4
k}{(2\pi )^4 }\gamma ^\nu  (\not p'-\not k + m)\gamma ^\mu  (\not
p-\not k + m)\gamma_\nu PP\frac{1}{k ^2-\mu^2 }
$$

$$
PP\left(\frac{m^2} {{(p'-k)^2 - m^2 }} \right)PP\left(\frac{m^2}
{{(p-k)^2  - m^2 }} \right)=F_1^{kr} (q^2 )\gamma ^\mu   +
\frac{{i\sigma ^{\mu \nu } q_\nu  }} {{2m}}F_2^{kr} (q^2 ).
$$

$F_2^{kr} (q^2 )$ in the two different method is the same and
$F_1^{kr} (q^2 )$ in the Krein regularization is:

$$
F_{1}^{kr} (q^2 )_{q^2  \to 0} = -\frac{e^2{q^2}} {16\pi^2
{m^2}}+\frac{3e^2{q^2}} {64\pi^2 {m^2}} -  \frac{{e^2 q^2 }}
{{12\pi^2 m^2 }}\left(\ln\frac{m} {\mu } - \frac{3} {8}\right) ,
$$

where $q^2=(p-p')^2$.

The singular terms of 3 standard graphs of QED  are replaced with
the two first terms in the resulted graphs in Krein space
quantization \cite{Dice2,forghan1}. By using the value of $F_1
(q^2 )$ and the photon self energy in Krein space, the value of
Lamb Shift is calculated to be 1018.19 MHz, whereas in standard
QED it is 1052.1 MHz; and its experimental value has been given as
1057.8 MHz. The small differences may be because of neglecting the
linear quantum gravitational effect and working in the one-loop
approximation \cite{forghan2}.

\section{Scalar Casimir effect for a sphere in de Sitter space}

  The Casimir force due to fluctuations of a free massless
  scalar field satisfying Dirichlet boundary conditions on a spherical
  shell in Minkowski space-time has been studied in \cite{mil}. Doing the calculations in Krein space, the two-point Green's
  function $G_K(x,t;x',t')$ is defined as the vacuum expectation
  value of the time-ordered product of two fields ($K$ and $T$ stand for
quantities in Krein space and the time ordered product,
respectively):
  \begin{equation}
  G_K(x,t;x',t')\equiv-i<0|T\Phi_K(x,t)\Phi_K(x',t')|0>.
  \label{1}
  \end{equation}

\begin{equation}
\Phi_K(x,t)=\sum_{\vec{k}} [( a_{\vec k}+ b_{\vec k}^\dag)
u_p(k,x) + ( a_{\vec k}^\dag+ b_{\vec k}) u_n(k,x)] \label{2}
\end{equation}

The operators $a^{\dag}(\vec k)$ and $a(\vec k)$ create and
destroy respectively the mode $u_p(k,x)$ with positive energy
$(k^0=\omega_{\vec k})$, which may be considered as the operators
of creation and annihilation of a particle and the operators
$b^{\dag}(\vec k)$ and $b(\vec k)$ create and destroy respectively
the mode $u_n(k,x)$ with negative energy $(-k^0=-\omega_{\vec
k})$, which may be considered as the operators of creation and
annihilation of an ``anti-particle'' in the inverse time
direction. The two sets of modes do not affect on each other and
in the standard QFT, the negative energy states are eliminated in
the quantum field operators, which is the origin of the appearance
of divergence. On the contrary, the divergence disappears by
taking these states into account.

  The two point Green function has to satisfy the Dirichlet boundary conditions on the
  shell:

  \begin{equation}
  G_K(x,t;x',t')|_{|x|=a}=0,
  \label{3}
  \end{equation}

  where $a$ is radius of the spherical shell. The stress-energy
  tensor in Krein space $T_K^{\mu\nu}(x,t)$ is given by

  \begin{equation}
  T_K^{\mu\nu}(x,t)\equiv\partial^{\mu} \Phi_K(x,t)\partial^{\nu} \Phi_K(x,t)-
  \frac{1}{2}\eta^{\mu\nu} \partial_{\lambda} \Phi_K(x,t)\partial^{\lambda}
  \Phi_K(x,t).
  \label{4}
  \end{equation}

  The radial Casimir force per unit area $\frac{F}{A}$ on the
  sphere, called Casimir stress, is obtained from the radial-radial component of the vacuum
  expectation value of the stress-energy tensor:

  \begin{equation}
  \frac{F}{A}=\langle0|T^{rr}_{in}-T^{rr}_{out}|0\rangle|_{r=a}.
  \label{5}
  \end{equation}

  Taking into account the relation (1) between the vacuum expectation value
  of the stress-energy tensor $T_K^{\mu\nu}(x,t)$ and the Green's
  function at equal times $ G_K(x,t;x',t)$ we obtain

  \begin{equation}
  \frac{F}{A}=\frac{i}{2}[\frac{\partial}{\partial r}\frac{\partial}{\partial
  r'}G_K(x,t;x',t)_{in}-\frac{\partial}{\partial r}\frac{\partial}{\partial
  r'}G_K(x,t;x',t)_{out}]|_{x=x',|x|=a}.
  \label{6}
  \end{equation}

  One may use of the above flat space calculation in de Sitter space-time by taking the de Sitter metric in conformally flat
form

  \begin{equation}
  ds^{2}=\Omega(\eta)[d\eta^{2}-\sum_{\imath=1}^{3}(dx^{\imath})^{2}],
  \label{7}
  \end{equation}

  where $\Omega(\eta)=\frac{\alpha}{\eta}$ and $\eta$ is the conformal time

  \begin{equation}
  -\infty <  \eta < 0.
  \label{8}
  \end{equation}

  Assuming a canonical quantization of the scalar field in Krein space, the conformally transformed quantized
   scalar field in de Sitter spacetime is given by

\begin{equation}
  \bar\Phi_K(x,\eta)=\sum_{k}[( a_{\vec k}+ b_{\vec k}^\dag)
  \bar u_{k}(\eta,x)+( a_{\vec k}^\dag+ b_{\vec k})\bar u_{k}^{\ast}(\eta,x)],
  \label{9}
  \end{equation}

  where $a_{k}^{\dagger}$ and
  $a_{k}$ are  creation and annihilation operators respectively
  and  the vacuum states associated with the physical modes $\bar u_{k}$
defined by
  $a_{k}|\bar 0\rangle=0 $, are called conformal vacuum.
  Given the flat space Green's function(1), we obtain

  \begin{equation}
 \bar G_K=-i\langle\bar{0}|T \bar \Phi_K(x,\eta)\bar
 \Phi_K(x',\eta^{'})|\bar{0}\rangle=\Omega^{-1}(\eta)\Omega^{-1}(\eta^{'})G_K,
 \label{10}
 \end{equation}

where $\bar\Phi_K(x,\eta)=\Omega^{-1}(\eta)\Phi_K(x,\eta)$ has
been used. Therefore, using Eqs.(\ref{5}), (\ref{6}) and
(\ref{10}) we obtain the total stress on the sphere in de Sitter
spacetime and using Krein space quantization as

\begin{equation}
\frac{\bar F}{A}=\frac{\eta^{2}}{\alpha^{2}} \frac{F}{A}.
\label{11}
\end{equation}

in accordance with the standard result \cite{Casimir2}.

\section{Conclusion}

In this paper we had a brief review on Krein space quantization.
For a first example, it was deduced that the 3 main divergent
graphs of standard QED are automatically regularized in Krein
space and the values of magnetic anomaly and Lamb shift in the
one-loop approximation are identical to the corresponding results
in standard calculations \cite{forghan2}. Due to the appearance of
negative norm states this method is similar to the Pauli-Villars
regularization, so it could be considered as a new method of
regularization called ''Krein regularization''.

For a second example, it was shown that using Krein space
quantization method of Scalar Casimir effect for a sphere in de
Sitter space leads to the same standard result. This method could
be easily generalized to non-Abelian gauge theory and quantum
gravity in the background field method, and could be used as an
alternative way for solving the non-renormalizability of quantum
gravity in the linear approximation.

Consequently for QED, the Krein space calculations just eliminates
the singularity in the theory without changing the physical
contents, and may provide an answer to the Feynman reply: ``A
Nobel prize for hiding the rushes under the carpet?".

\vskip 0.5 cm

\noindent {\bf{Acknowledgments}}: The authors would like to thank
M. V. Takook for useful discussions.


\begin{thebibliography}{[1]}


\bibitem{anilto} I. Antoniadis, J. Iliopoulos, T.N. Tomaras, Nuclear
Phys. B, {\bf 462}, 437 (1996).

\bibitem{gagarota} T. Garidi et al, J. Math. Phys., {\bf 49}, 032501 (2008);
T. Garidi et al, J. Math. Phys., {\bf 44}, 3838 (2003); S.
Behroozi et al, Phys. Rev. D, {\bf 74}, 124014 (2006).

\bibitem{derotata} M. Dehghani et al, Phys. Rev. D, {\bf 77}, 064028 (2008); M.V. Takook et al,
J. Math Phys., {\bf 51}, 032503 (2010).

\bibitem{al} B. Allen, Phys. Rev. D, {\bf 32}, 3136 (1985).

\bibitem{gareta} J.P. Gazeau, J. Renaud, M.V. Takook, Class. Quan.
Grav., {\bf 17}, 1415 (2000), gr-qc/9904023.

\bibitem{ta3} M.V. Takook, Mod. Phys. Lett. A, {\bf 16}, 1691 (2001),
gr-qc/0005020.

\bibitem{for2} H.L. Ford, Quantum Field Theory in Curved Spacetime,
gr-qc/9707062.

\bibitem{ta4} M.V. Takook, Int. J. Mod. Phys. E, {\bf 11}, 509 (2002),
gr-qc/0006019.

\bibitem{rota} S. Rouhani, M.V. Takook, Int. J. Theor. Phys., {\bf 48}, 2740 (2009).


\bibitem{Boson} F. Payandeh, M. Mehrafarin, S. Rouhani, M. V. Takook,
UJP, {\bf 53}, 1203 (2008).

\bibitem{AIP} F. Payandeh, M. Mehrafarin, M. V. Takook,
AIP Conf Proc, {\bf 957}, 249 (2007).

\bibitem{QFT} F. Payandeh, Journal of Physics: Conference Series {\bf 174}, 012056
(2009).

\bibitem{Moller} F. Payandeh, M. Mehrafarin, M. V. Takook, Science in China Series G: Physics, Mechanics and
Astronomy, {\bf 52}, 212 (2009).

\bibitem{Takook} S. Rouhani, M.V. Takook, Int. J. Theor. Phys {\bf 48}, 2740 (2009) arXiv:gr-qc/0607027v1.

\bibitem{Setare} M. R. Setare, F. Darabi, Int. J. Mod. Phys. D {\bf 17}, 2261 (2008) arXiv:hep-th/0511077v1.

\bibitem{Casi48} H. B. G. Casimir, Proc. K. Ned. Akad. Wet. {\bf
51}, 793 (1948).

\bibitem{Naseri} H. Khosravi, M. Naseri, S. Rouhani and M. V.
Takook, Phys. Lett. B {\bf 640}, 48 (2006).

\bibitem{Casimir2} F. Payandeh, ISRN High Energy Physics, {\bf
2012} 714823 (2012).


\bibitem{Takook2} A. Refaei , M. V. Takook, Mod. Phys. lett. A, {\bf
26} 31 (2011).


\bibitem{ba} Barth N.H., Christensen S.M., Phys. Rev. D, 28(1983)1876.
\bibitem{ho} Horva P., Phys. Rev. D, 79(2009)084008, arXiv:0901.3775.
\bibitem{ka} Kaku M., Oxford University Press, (1993), {\it  QUANTUM FIELD THEORY :A MODERN INTRODUCTION}.



\bibitem{pesc} E. Peskin, D.V. Schroeder, Perseus Books, $(1995)$, An Introduction  in Quantum Field Theory.


\bibitem{Dice2} F. Payandeh, Journal of Physics: Conference Series {\bf 306}
012054 (2011).

\bibitem{forghan1} A. Zarei, M.V. Takook, B. Forghan, INT. J. Theor.
Phys. {\bf 50} 2460 (2011).

\bibitem{forghan2} B. Forghan, M.V. Takook, A. Zarei, Krein
Regularization of QED, in preparation.

\bibitem{mil}C. M. Bender and K. A. Milton, Phys. Rev. {\bf D55},
6547 (1994).





\end{thebibliography}
\end{document}